\pdfoutput=1
\documentclass{article}

\PassOptionsToPackage{numbers, compress}{natbib}

\usepackage[dblblindworkshop, final]{neurips_2026} %
\workshoptitle{Sim2Science: ML with Imperfect Scientific Models}

\usepackage[utf8]{inputenc} %
\usepackage[T1]{fontenc}    %
\usepackage{hyperref}       %
\usepackage{url}            %
\usepackage{booktabs}       %
\usepackage{amsfonts}       %
\usepackage{nicefrac}       %
\usepackage{microtype}      %
\usepackage[table]{xcolor}  %

\definecolor{exptregion}{gray}{0.92} %
\usepackage{amsmath}            %
\usepackage{amssymb}            %
\usepackage{graphicx}
\usepackage{siunitx}
\usepackage[version=4]{mhchem}  %
\usepackage{multirow}
\usepackage{makecell}
\usepackage{pifont}             %
\graphicspath{{figures/}}
\DeclareSIUnit\atom{atom}
\DeclareSIUnit\angstrom{\text{\AA}}  %

\newcommand{\cm}{\textcolor{green!45!black}{\ding{51}}}
\newcommand{\xm}{\textcolor{red!70!black}{\ding{55}}}
\newcommand{\rep}[2]{{\scriptsize\selectfont
  \ifnum#1>0 \cm\fi\ifnum#1>1 \cm\fi\ifnum#1>2 \cm\fi\ifnum#1>3 \cm\fi\ifnum#1>4 \cm\fi
  \ifnum#2>0 \xm\fi\ifnum#2>1 \xm\fi\ifnum#2>2 \xm\fi\ifnum#2>3 \xm\fi\ifnum#2>4 \xm\fi}}

\title{
Challenging Atomistic Foundation Models on Disordered Carbons
}

\author{%
  Joseph Hart\textsuperscript{1} \quad
  Samuel G.\ H.\ Brookes\textsuperscript{1} \quad
  Fabian L. Thiemann\textsuperscript{2} \quad
  Christoph Schran\textsuperscript{1} \\
  \textsuperscript{1}Cavendish Laboratory, Department of Physics \\
  University of Cambridge \\
  \textsuperscript{2}Microsoft Research AI for Science \\
}

\begin{document}

\maketitle

\begin{abstract}
    Foundation machine-learned interatomic potentials (MLIPs) promise broad transferability, yet their out-of-domain reliability is hard to establish. 
    Here, we use pure and O/H-containing disordered carbon spanning multiple phases, competing hybridisations and porous frameworks to stress-test six popular foundation MLIPs. 
    We first screen all six on density functional theory (DFT) energy and force errors and binding curves, then subject four to melt-quench-anneal molecular dynamics and evaluate the resulting structural observables.
    Binding curves and O/H-containing trajectories reveal unphysical extrapolation and dynamical failures that low DFT energy and force errors fail to reveal.
    Of the models tested, MACE-MH-1 is the most consistently robust, with low DFT errors and stable MD.
    Despite differing architectures, OMAT-based models share structural biases, highlighting the importance of training coverage.
    All four models generate structures of nanoporous carbon that reproduce the principal experimental and specialised carbon potential signatures. 
    Incorporating O/H leaves short-range order largely unchanged but shifts pore-size distributions towards larger pores. 
    These results emphasise the need for application-specific validation and extend molecular dynamics modelling of nanoporous carbon towards experimentally realistic structures.
\end{abstract}

\section{Introduction}

Machine-learned interatomic potentials (MLIPs) approximate the potential energy surface (PES) typically at DFT level, enabling molecular dynamics at far larger length and time scales than direct DFT calculations~\cite{jacobs2025practical,wang2024machine}.
Conventional MLIPs are generally system-specific, requiring new reference calculations and training different elements, phases or bonding environments. 
Foundation MLIPs instead learn from broad datasets and can be deployed directly or adapted with comparatively little additional data~\cite{kaur2025data}.
Early models such as M3GNet~\cite{chen2022universal}, CHGNet~\cite{deng2023chgnet} and MACE-MP-0~\cite{batatia2025foundation} focused largely on inorganic crystals, whereas recent models include surfaces, molecules and broad thermodynamic conditions~\cite{batatia2025cross,kim2026optimizing,li2606dpa4,wood2025family,kabylda2025molecular,rhodes2025orb,batatia2026mace,mazitov2025pet}.
However, broad training coverage does not ensure out-of-distribution reliability.
Near-equilibrium training data can systematically soften the learned PES~\cite{deng2025systematic}, while energy and force errors can miss spurious minima that destabilise dynamics or bias observables~\cite{fu2022forces}.
Recent benchmark efforts therefore supplement energy and force errors with property- and simulation-based tests~\cite{riebesell2023matbench,chiang2025mlip,jakob2025universally}. 
However, selecting a foundation MLIP for a specific out-of-domain workflow remains challenging.
Nanoporous carbon provides a stringent and technologically relevant test. Its formation spans liquid, amorphous and graphitic environments over extreme temperatures and long time scales, while its structure combines sp, sp$^2$ and sp$^3$ bonding, under-coordinated surfaces, defects and extended pore networks. Nanoporous carbon is also particularly relevant in energy storage, carbon capture and water treatment, making its accurate atomistic description an important goal~\cite{liuHighlyDisorderedNanoporous2025,corrente_slit_2024,simon2010charge,porada2013review}.
These challenges have motivated substantial effort towards specialised carbon potentials such as GAP-17, GAP-20 and GAP-22~\cite{deringer2017machine,rowe2020accurate,wang2022structure}. Validation is nevertheless difficult because local bonding and pore-scale morphology require observables spanning several length scales.
A foundation MLIP that remains reliable across these environments would demonstrate meaningful transfer and provide a robust starting point for targeted fine-tuning~\cite{chorna2026comparing,tompa2026fine}.
Existing melt-quench models of nanoporous carbon have largely treated the material as pure carbon~\cite{wang2022structure,de2016graphitization,de2018carbide,de2019transferability}. 
Extending specialised potentials to O/H-containing compositions requires new reference data and training, whereas foundation MLIPs offer a route to chemical realism and to probing heteroatom effects on framework formation.
Here, we evaluate foundation MLIPs through external DFT benchmarks, two- and many-body binding curves, melt-quench-anneal stability and structural observables. Low DFT energy and force errors do not guarantee physical extrapolation or stable dynamics, and OMAT-based models exhibit common structural biases despite differing architectures. The strongest models reproduce special-purpose MLIPs and experimental references. Although O/H incorporation has little effect on local carbon order, it promotes a more open pore network, revealing a role for heteroatoms in framework assembly.

\section{Simple benchmarks reveal complementary model failures}
\label{sec:benchmarking}

\paragraph{Community DFT datasets provide an initial screen.}

\begin{table}[b]
  \caption{\textbf{Energy and force errors on PBE carbon datasets.} Energy and force mean absolute errors (MAEs), relative force errors and a normalised aggregate score are reported. Higher scores are better. Darker shading marks larger errors. Bold and underlined values mark the two lowest MAEs in each column. Scoring is detailed in Appendix~\ref{app:energies_and_forces}.}
  \label{tab:errors}
  \centering
  \footnotesize
  \setlength{\tabcolsep}{2.2pt}
  \begin{tabular}{l c ccc ccc ccc ccc ccc}
    \toprule
    \multirow{3}{*}{{\scriptsize\makecell[l]{Foundation MLIP\\[1pt]$E$ / meV\,atom$^{-1}$\\[-2pt]$F$ / meV\,\AA$^{-1}$\\[-2pt]$F_{\mathrm{rel}}$ / \%}}} & \multirow{3}{*}{Score} & \multicolumn{12}{c}{Diverse carbon} & \multicolumn{3}{c}{\makecell{Graphene\\oxide}} \\
    \cmidrule(lr){3-14}\cmidrule(lr){15-17}
     & & \multicolumn{3}{c}{sp$^2$} & \multicolumn{3}{c}{sp$^3$} & \multicolumn{3}{c}{Amor./liq.} & \multicolumn{3}{c}{Clusters} & \multicolumn{3}{c}{} \\
    \cmidrule(lr){3-5}\cmidrule(lr){6-8}\cmidrule(lr){9-11}\cmidrule(lr){12-14}
     & & $E$ & $F$ & $F_{\mathrm{rel}}$ & $E$ & $F$ & $F_{\mathrm{rel}}$ & $E$ & $F$ & $F_{\mathrm{rel}}$ & $E$ & $F$ & $F_{\mathrm{rel}}$ & $E$ & $F$ & $F_{\mathrm{rel}}$ \\
    \midrule
    ORB-v3-consv-inf-OMAT & \textbf{0.764} & \underline{31} & \underline{190} & 11 & \textbf{87} & 162 & \cellcolor{blue!8}15 & \underline{17} & \underline{249} & \cellcolor{blue!8}15 & \cellcolor{blue!8}364 & \textbf{452} & \cellcolor{blue!8}18 & \cellcolor{blue!8}62 & 189 & 14 \\
    MACE-MH-1-OMAT & \underline{0.756} & 35 & 231 & 13 & 122 & \underline{161} & \cellcolor{blue!8}15 & \cellcolor{blue!8}22 & \cellcolor{blue!8}280 & \cellcolor{blue!8}17 & \cellcolor{blue!8}475 & \underline{477} & \cellcolor{blue!8}19 & \cellcolor{blue!8}46 & \underline{172} & 13 \\
    MatterSim-5M & 0.705 & \textbf{28} & \cellcolor{blue!8}241 & 13 & 109 & \textbf{136} & 13 & \cellcolor{blue!8}24 & \cellcolor{blue!8}342 & \cellcolor{blue!8}21 & \cellcolor{blue!8}580 & 552 & \cellcolor{blue!8}22 & \cellcolor{blue!8}59 & \cellcolor{blue!8}348 & \cellcolor{blue!16}27 \\
    MACE-MP-0b3 & 0.650 & \cellcolor{blue!16}87 & \cellcolor{blue!16}493 & \cellcolor{blue!16}27 & 117 & \cellcolor{blue!8}304 & \cellcolor{blue!16}29 & \cellcolor{blue!30}139 & \cellcolor{blue!16}565 & \cellcolor{blue!16}34 & \underline{294} & \cellcolor{blue!8}809 & \cellcolor{blue!16}32 & \textbf{26} & \cellcolor{blue!16}428 & \cellcolor{blue!16}33 \\
    MACE-OMAT-0 & 0.648 & \cellcolor{blue!16}139 & \cellcolor{blue!16}1191 & \cellcolor{blue!30}66 & \underline{102} & \cellcolor{blue!8}259 & \cellcolor{blue!8}24 & \cellcolor{blue!16}54 & \cellcolor{blue!8}414 & \cellcolor{blue!16}25 & \textbf{233} & 645 & \cellcolor{blue!16}25 & \underline{34} & \cellcolor{blue!8}336 & \cellcolor{blue!16}26 \\
    UMA-S-1.1-OMAT & 0.520 & \cellcolor{blue!8}64 & \textbf{156} & 9 & \cellcolor{blue!16}827 & \cellcolor{blue!16}440 & \cellcolor{blue!16}41 & \textbf{12} & \textbf{168} & 10 & \cellcolor{blue!30}3015 & \cellcolor{blue!30}10187 & \cellcolor{blue!45}401 & \cellcolor{blue!16}215 & \textbf{127} & 10 \\
    \bottomrule
  \end{tabular}
\end{table}

Energy and force errors on unseen DFT datasets provide an efficient initial screening of transferability. We use two complementary datasets calculated using the Perdew–Burke–Ernzerhof (PBE) functional, consistent with the model's training set. The first spans more than 10,000 pure-carbon configurations covering sp$^2$ and sp$^3$ bonding, crystalline, amorphous and liquid phases, and isolated clusters~\cite{qamar2023atomic}. The second contains more than 3,000 O/H-functionalised graphene-oxide structures~\cite{el2024accelerated}. Together they probe both the carbon framework's structural diversity and its heteroatom chemistry, where a suitable model should perform well on both.
ORB-v3 achieves the highest overall score in Table~\ref{tab:errors}, followed closely by MACE-MH-1. Since absolute force errors depend on the underlying force scale, particularly for high temperature configurations, we also report the relative force error, $F_{\mathrm{rel}}$~\cite{ying2025advances}. ORB-v3 and MACE-MH-1 are the only models below 20\% across all environments. MatterSim exceeds 20\% for amorphous/liquid carbon, clusters and graphene oxide, while UMA performs well for several periodic environments but struggles with sp$^3$ carbon and isolated clusters. The latter failure is not confined to a few outliers, with a substantial fraction of cluster configurations exceeding 10,000~meV\,\AA$^{-1}$ (Figure~\ref{fig:force-violins}a).
Although both UMA and MACE-MH-1 are exposed to molecular and cluster configurations through the OMOL dataset~\cite{levine2025open}, only MACE-MH-1 performs consistently across the tested domains. 
This difference may reflect model design. The MACE-MH-1 OMAT head targets broad transfer, whereas UMA offers task-specific heads, which is challenging for simulations spanning several domains.
Overall, these benchmarks provide an initial screen, but low energy and force errors alone do not ensure physical behaviour, stable dynamics or accurate downstream observables~\cite{fu2022forces}.

\paragraph{Binding curves reveal unphysical extrapolation.}

\begin{figure}[t]
  \centering
  \includegraphics[width=0.9\linewidth]{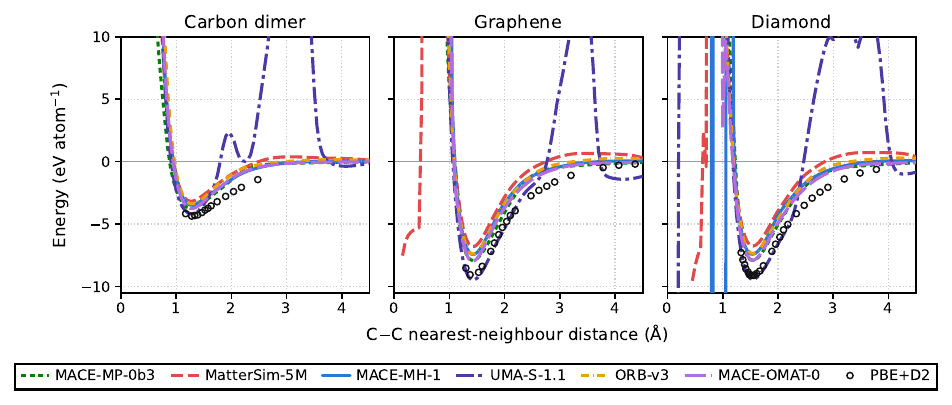}
    \caption{\textbf{Two- and many-body binding curves.} Energy per atom against nearest-neighbour C--C distance for the carbon dimer and isotropically scaled graphene and diamond, referenced to each model's dissociation limit. D2 dispersion is applied to match the PBE+D2 reference~\cite{qamar2023atomic}. D3-corrected curves are shown in Figure~\ref{fig:binding_D3}.}
  \label{fig:binding_D2}
\end{figure}

The interaction between two atoms is the simplest test of a potential's physical behaviour.
We first examine this limit using the carbon-dimer binding curve in Figure~\ref{fig:binding_D2}, which probes both bond formation and short-range repulsion.
Following Qamar et al.~\cite{qamar2023atomic}, we then extend this by isotropically scaling graphene and diamond cells across wide volume ranges to probe many-body behaviour.
Although this extreme compression at short distances is not directly relevant to nanoporous carbon generation, it can expose PES ``holes'' that unphysically stabilise compressed, over-coordinated structures. Large barriers make sampling these regions unlikely in ambient-pressure MD, so we treat them as extrapolation diagnostics rather than exclusion criteria~\cite{batatia2025foundation}.
All models reproduce the PBE+D2 equilibrium distance and dimer repulsion. With the exception of UMA, however, they underestimate the well depth and therefore underbind. UMA closely matches the reference near equilibrium but develops two additional dimer minima at 2.2 and 4.1~\AA, separated by a barrier exceeding 10~eV\,atom$^{-1}$ (Figure~\ref{fig:binding_log}). Expanded graphene and diamond show analogous minima. Despite low sp$^2$ errors in Table~\ref{tab:errors}, UMA does not reliably extrapolate far from equilibrium here.
MatterSim, MACE-MH-1 and UMA exhibit PES holes under strong diamond compression, with MatterSim also failing for graphene. MACE-OMAT-0 shows an unphysical decrease in the repulsive diamond curve but recovers before a hole forms.
All MLIPs nonetheless reproduce the dimer repulsion correctly, indicating that the holes are a many-body effect, where even an analytical ZBL pair repulsion term does not guarantee a physical many-body PES.

\section{Dynamical validation of nanoporous carbon generation}
\label{sec:observables}

\paragraph{Melt-quench MD stability.}

\begin{table}[b]
  \caption{\textbf{Stability of O/H-containing melt-quench-anneal simulations.} Each entry shows the outcomes of five independent runs at the indicated density, where \cm{} denotes completion and \xm{} denotes a simulation failure. The shaded region indicates the experimentally relevant density range for supercapacitor materials. All corresponding pure-carbon simulations completed successfully.}
  \label{tab:stability}
  \centering
  \footnotesize
  \setlength{\tabcolsep}{4pt}
  \begin{tabular}{
    l
    *{3}{>{\columncolor{exptregion}}c}
    *{4}{c}
  }
    \toprule
    \multirow{2}{*}{Foundation MLIP}
      & \multicolumn{7}{c}{Density / \si{\gram\per\cubic\centi\metre}} \\
    \cmidrule(lr){2-8}
      & 0.5 & 1.0 & 1.5 & 2.0 & 2.5 & 3.0 & 3.5 \\
    \midrule
    ORB-v3-consv-inf-OMAT
      & \rep{4}{1} & \rep{5}{0} & \rep{4}{1}
      & \rep{5}{0} & \rep{5}{0} & \rep{5}{0} & \rep{5}{0} \\
    MACE-MP-0b3
      & \rep{5}{0} & \rep{5}{0} & \rep{5}{0}
      & \rep{5}{0} & \rep{5}{0} & \rep{0}{5} & \rep{0}{5} \\
    MACE-OMAT-0
      & \rep{5}{0} & \rep{5}{0} & \rep{3}{2}
      & \rep{5}{0} & \rep{0}{5} & \rep{0}{5} & \rep{0}{5} \\
    MACE-MH-1-OMAT
      & \rep{5}{0} & \rep{5}{0} & \rep{5}{0}
      & \rep{5}{0} & \rep{5}{0} & \rep{5}{0} & \rep{5}{0} \\
    \bottomrule
  \end{tabular}
\end{table}

Simulation stability across the intended configurational space provides a more demanding assessment of out-of-domain suitability. Although stability does not guarantee accurate observables, it is necessary for practical use. The melt-quench-anneal protocol samples out-of-equilibrium configurations across extreme temperatures, phases and bonding environments in independent simulations spanning seven fixed densities.
Previous melt-quench studies of nanoporous carbon have largely been restricted to pure-carbon systems. We therefore simulate both pure carbon and O/H-containing compositions representative of commercial supercapacitor carbons, simultaneously testing chemical transfer and enabling the generation of more realistic electrode structures.
For each model and composition, five independent 216-atom simulations were performed at seven densities between 0.5 and 3.5~g\,cm$^{-3}$ using different initialisation seeds, including the experimentally relevant range of 0.5--1.5~g\,cm$^{-3}$ for supercapacitor carbons~\cite{kim2016hierarchically,jovanovic2025operando,peng2026realistic}. Full details are provided in Appendix~\ref{app:protocol}.
Each model required 70 trajectories totalling 29.2~ns, making testing all six impractical. UMA was excluded after its binding-curve failures, while MatterSim was omitted to limit cost. The remaining four were tractable using \texttt{symmetrix} for MACE~\cite{kovacs2025mace,batatia2025foundation} and \texttt{AlchemiDFTD3} for ORB's D3 correction.
Every pure-carbon simulation completed successfully for all four models and densities, whereas the O/H-containing systems clearly discriminate between models (Table~\ref{tab:stability}).
MACE-MH-1 is the only model to complete all 35 simulations. ORB-v3 completes 33, with isolated failures at 0.5 and 1.5~g\,cm$^{-3}$. Although these could reflect unfavourable initial configurations, both occur during annealing and may therefore be associated with the evolution of the O/H-containing framework and the formation of gas molecules. MACE-OMAT-0 is the least stable model tested, failing in 17 simulations, including two at 1.5~g\,cm$^{-3}$ and every run above 2.0~g\,cm$^{-3}$. Its temporary recovery at 2.0~g\,cm$^{-3}$ suggests that stability may not be governed by density alone.
MACE-MP-0b3 remains stable up to 2.5~g\,cm$^{-3}$ but fails in every simulation at 3.0 and 3.5~g\,cm$^{-3}$, where the structures become increasingly sp$^3$-dominated. This behaviour is not anticipated by the external DFT benchmarks, for which its relative force errors for sp$^2$ and sp$^3$ carbon are similar at 27\% and 29\%, respectively. Energy and force errors within individual domains therefore do not necessarily predict stability when a trajectory samples the phase space connecting them.

\paragraph{Structural validation against specialised potentials and experiment.}

\begin{figure}[t]
  \centering
  \includegraphics[width=0.94\linewidth]{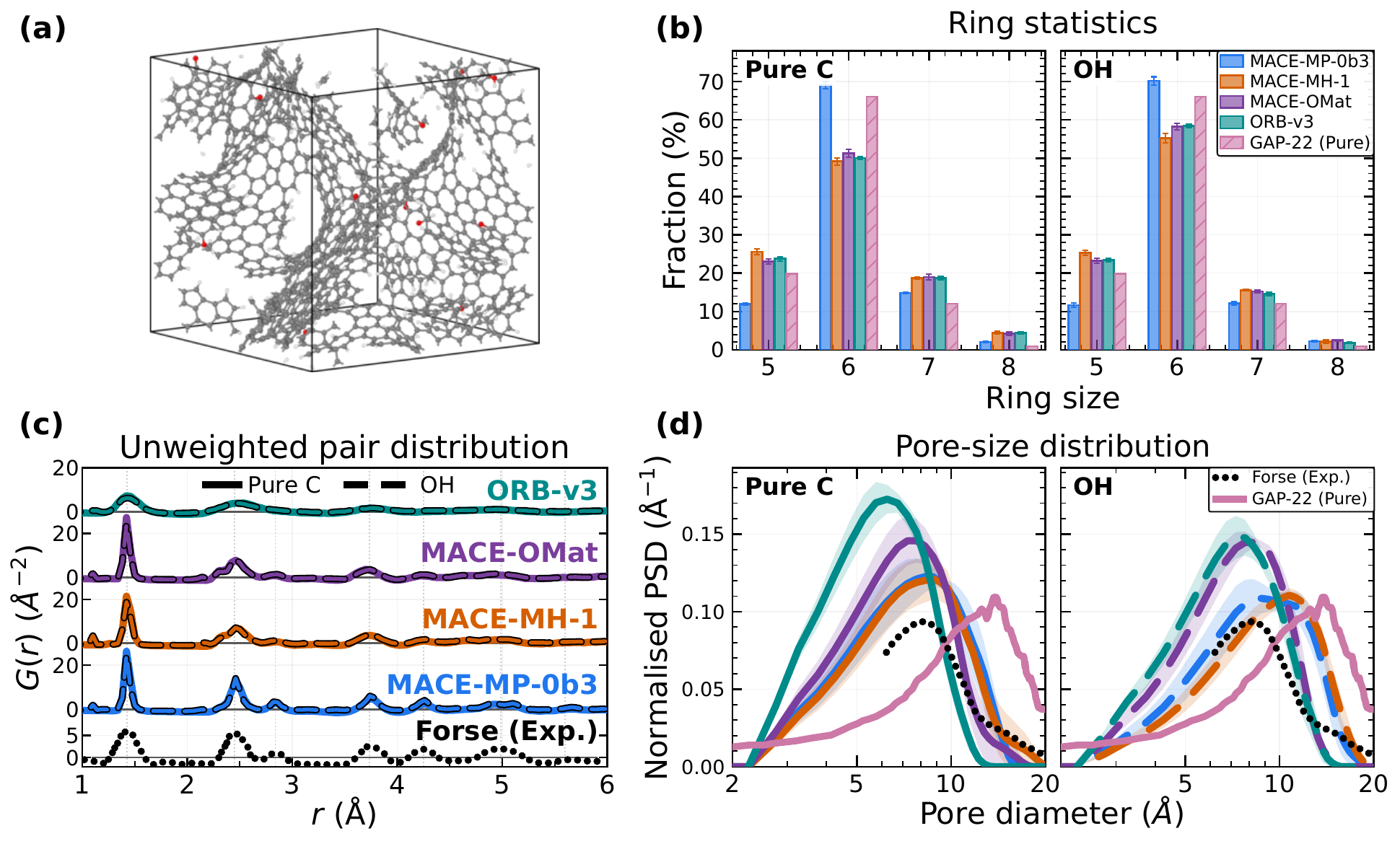}
    \caption{\textbf{Structural validation of nanoporous carbons.} (a) Representative O/H-containing structure generated by MACE-MH-1 at 1.0~g\,cm$^{-3}$. (b) Ring-size distributions against pure-carbon GAP-22 ($\rho = 0.9$~g\,cm$^{-3}$)~\cite{wang2022structure}. (c) Pair distribution functions and (d) pore-size distributions, with experimental references from Ref.~\cite{forse2015new}. Curves show means over five seeds and shading or error bars indicate the standard error. Solid and dashed lines denote pure and O/H-containing carbon.}
  \label{fig:validation}
\end{figure}

Having established simulation stability, we assess whether the generated structures reproduce computational and experimental observables across local bonding, network topology and porosity.
Five independent 1728-atom structures were generated for each model and composition at 1.0~g\,cm$^{-3}$, with a representative MACE-MH-1 structure shown in Figure~\ref{fig:validation}a.
Figure~\ref{fig:validation}b compares the ring-size distributions with GAP-22, a general carbon model validated for amorphous and nanoporous carbon~\cite{wang2022structure}. ORB-v3, MACE-OMAT-0 and MACE-MH-1, all trained on OMAT, yield nearly identical distributions for pure carbon, reproducing the five-membered-ring population but underpredicting six-membered rings. MACE-MP-0b3 instead underpredicts five-membered rings while reproducing the six- and seven-membered populations.
The common behaviour of the three OMAT-based models, despite their different architectures, points to their shared training domain, which contains high temperature AIMD trajectories.
These models retain more under-coordinated carbon in the amorphous precursor
(Figure~\ref{fig:app_coord}). Local bond closure into non-hexagonal rings can alleviate these sites, resulting in curvature and lower graphitic order, whereas forming extended six-membered domains requires more collective rearrangement. With O/H, the difference between models is smaller, consistent with heteroatoms terminating under-coordinated sites and reducing the need for non-hexagonal ring formation.
All models reproduce the principal experimental pair distribution function (PDF) peak positions (Figure~\ref{fig:validation}c), although the OMAT-based models produce broader features, particularly ORB-v3, supporting the lower graphitic order inferred from their ring statistics.
The higher first-peak amplitudes of the MACE PDFs are consistent with previous simulations~\cite{front2026density,deringer2018towards}. Applying the experimental cutoff, $Q_{\max}=24$~\AA$^{-1}$, broadens the simulated peaks and aligns their amplitudes with experiment, but suppresses structural detail~\cite{de2017structural}. We therefore show the un-broadened simulated curves. Pure and O/H-containing PDFs are otherwise similar, indicating that the heteroatoms have little effect on short-range carbon structure.
Figure~\ref{fig:validation}d compares pore-size distributions (PSDs) calculated using \texttt{PoreTorch} with experimental Ti-CDC-1000 data~\cite{forse2015new} and GAP-22~\cite{wang2022structure}.
As experimental PSDs depend on the adsorption model, we interpret the comparison semi-quantitatively~\cite{neimark2009quenched}. All foundation models reproduce the characteristic experimental pore-size range, with the closest agreement generally obtained for the O/H-containing structures. MACE-MP-0b3 and MACE-MH-1 agree the most closely with GAP-22, whereas ORB-v3 and MACE-OMAT-0 favour smaller pores. The larger simulated peak amplitudes are partly attributable to finite-size effects, although the characteristic pore sizes are adequately converged (Figure~\ref{fig:app_PSD}). GAP-22 predicts larger pores, consistent with its lower density of 0.9~g\,cm$^{-3}$ and substantially larger 130,000-atom cells.
Across all models, incorporating O and H shifts the PSDs towards larger pores while leaving the PDFs largely unchanged. This shift partly reflects gas formation, which lowers the framework density, and O/H termination of under-coordinated carbon, which suppresses ring closure and promotes a more open framework.
Heteroatoms therefore have little influence on short-range carbon order but alter framework assembly at longer length scales. As pore morphology controls accessible surface area and ion transport, pure-carbon models may miss structurally important effects of experimentally relevant compositions.

\section{Conclusion}
\label{sec:conclusion}

Out-of-domain transfer is a multifaceted challenge that cannot be assessed through a single error metric. For carbon, low DFT energy and force errors, even on out-of-equilibrium structures, do not guarantee physically reasonable extrapolation or stable dynamics. 
Binding curves and melt-quench-anneal simulations expose failures missed by conventional benchmarks, demonstrating the value of progressively testing static accuracy, physicality, dynamical stability and downstream observables.
Across this hierarchy of tests, MACE-MH-1 provides the most consistent performance among the models evaluated.
The carbon tests are effective discriminators because they combine inexpensive out-of-domain screening, stringent MD stability testing and validation against physical observables derived from predicted energies and forces. 
This complements broader benchmarking efforts such as ML-PEG~\cite{ml-peg}.
We therefore suggest their inclusion in standard foundation-MLIP test suites, extending this evaluation to models not examined here, such as PET-MAD~\cite{mazitov2025pet} and DPA4~\cite{li2606dpa4}.
Although computationally expensive, applying MD benchmarks before production simulations can avoid greater downstream cost and expose targets for model improvement.
Here, the validated models also enable melt-quench generation to be extended beyond the pure-carbon approximation. At an experimentally relevant density, ORB-v3, MACE-MP-0b3, MACE-OMAT-0 and MACE-MH-1 all reproduce the principal local and pore-scale features of specialised-potential and experimental references, although the OMAT-based models share structural biases despite differing architectures. Incorporating O/H leaves short-range order largely unchanged but shifts pore-size distributions towards larger pores, suggesting that the realistic composition affects longer-range morphology.

\paragraph{Limitations \& Outlook.} 

Stable dynamics and agreement with selected observables do not establish accuracy across the full potential energy surface or uniquely determine disordered-carbon structures.
Our dynamical tests use a single generation protocol, while structural validation is restricted to one density, one O/H composition and finite length scales.
Future work should utilise short \textit{ab initio} MD benchmarks and explore broader compositions and protocols to determine how O/H incorporation affects pore formation, capacitance and ion transport. The failure configurations identified here also provide targets for fine-tuning and general improvement of atomistic foundation models. Further work should also assess which DFT functionals most accurately describe the diverse bonding environments sampled here and identify any artefacts specific to the generalised-gradient approximation.

\begin{ack}

We would like to thank Professor Alexander Forse and the Forse group for valuable discussions on experimentally realistic nanoporous carbons.
S.G.H.B and C.S. acknowledge support from the Isaac Newton Trust G122390 and the UKRI reference EP/V062654/1.
J.H. and C.S. acknowledge support from EPSRC through an Industrial Doctoral Landscape Award (UKRI1993), co-sponsored by IBM.
We are grateful for computational support and resources from the UK national high-performance computing service, Advanced Research Computing High End Resource (ARCHER2).
Access for ARCHER2 were obtained via the UK Car-Parrinello consortium, funded by EPSRC grant reference EP/P022561/1, and via two Access to HPC calls.
The authors acknowledge the use of resources provided by the Isambard-AI National AI Research Resource (AIRR)~\cite{bristol-ai}.
Isambard-AI is operated by the University of
Bristol and is funded by the UK Government's Department for Science, Innovation and Technology (DSIT) via UK Research and Innovation; and the Science and Technology Facilities Council [\texttt{ST/AIRR/I-A-I/1023}].
The authors declare no competing interests.

\end{ack}

\clearpage

\appendix

\section{Energies and forces dataset, evaluation and scoring}
\label{app:energies_and_forces}

The benchmarks use two external PBE datasets. The diverse-carbon dataset~\cite{qamar2023atomic} is divided into sp$^2$, sp$^3$, amorphous/liquid, cluster and general-bulk subsets, while the graphene-oxide dataset~\cite{el2024accelerated} contains carbon, oxygen and hydrogen. Configurations under extreme lattice compression were excluded from the main comparison because their large errors dominate the means and obscure performance across the majority of each subset. The corresponding high-compression regime is assessed separately using the binding curves in Figure~\ref{fig:binding_D2}.

\begin{figure}[htbp]
  \centering
  \includegraphics[width=\linewidth]{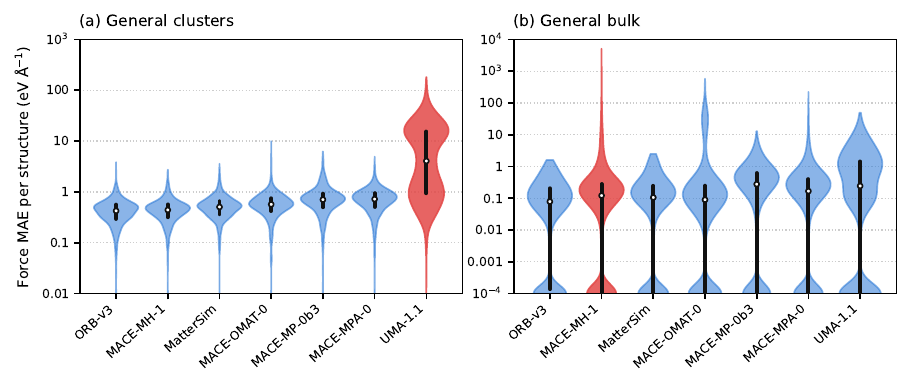}
    \caption{\textbf{Per-structure force-error distributions.}
    Force MAEs for (a) isolated-cluster and (b) general-bulk configurations.
    Violins show kernel-density estimates of the $\log_{10}$-transformed force
    MAEs, with white markers denoting medians and bars the interquartile ranges.
    The low-error population in (b) arises from 1414 structures whose reference
    forces vanish due to lattice symmetry.}
  \label{fig:force-violins}
\end{figure}

As discussed in the main text, the large UMA-S-1.1 cluster error is not caused by a small number of outliers. Table~\ref{tab:errors_clusters} and Figure~\ref{fig:force-violins}a show that its median force error is at least 5.6 times that of every other model and exceeds the maximum error recorded for any individual cluster by ORB-v3, MACE-MH-1 or MatterSim. The poor result therefore reflects a broad shift of the error distribution.
The general-bulk subset was excluded because it is the least directly relevant subset for nanoporous carbon and its force statistics are controlled by small groups of unusual structures. Of the 5300 structures, 1414 are high-symmetry lattice scans whose reference forces vanish at every lattice parameter, lowering the median while contributing almost nothing to the mean force error (Figure~\ref{fig:force-violins}b). At the opposite extreme, ten vacancy-containing, randomly displaced bcc and fcc cells account for approximately 72\% of the accumulated MACE-MH-1 error. Its median force error is therefore 122~meV\,\AA$^{-1}$, compared with a mean of 18\,998~meV\,\AA$^{-1}$. This large mean records a genuine failure on structures containing short interatomic distances and high coordination, but is not representative of performance across the subset. These interactions are isolated more directly by the binding curves in Figure~\ref{fig:binding_D2}. Including general bulk would place MatterSim second with a score of 0.717 and MACE-MH-1 third with 0.644. Full results are reported in Table~\ref{tab:errors_bulk} and Figure~\ref{fig:force-violins}a.

Energy errors are reported as MAEs per atom, while force errors are averaged over all Cartesian components. The overall score follows related benchmarking frameworks~\cite{ml-peg}. Each energy and force error is mapped linearly onto a dimensionless score between predefined ``good'' and ``bad'' thresholds. Errors at or below the good threshold receive a score of 1, while errors at or above the bad threshold receive 0. The overall score is the unweighted mean of the ten energy and force scores across the five main subsets. Relative force errors are not weighted in the score because they give the same within-subset ranking as the absolute force errors.

\begin{table}[htbp]
    \caption{\textbf{Energy and force errors for isolated clusters.}
    MAEs and per-structure force-error statistics over the 2306 cluster structures from the diverse carbon dataset~\cite{qamar2023atomic}. Energy and force units are meV\,atom$^{-1}$ and meV\,\AA$^{-1}$, respectively. Bold and underlined
    values mark the best and second-best results.}
  \label{tab:errors_clusters}
  \centering
  \footnotesize
  \setlength{\tabcolsep}{5pt}
  \begin{tabular}{l ccc cc}
    \toprule
    \multirow{2}{*}{Foundation MLIP} & \multicolumn{3}{c}{MAE}
      & \multicolumn{2}{c}{Per-structure force MAE} \\
    \cmidrule(lr){2-4}\cmidrule(lr){5-6}
     & $E$ & $F$ & $F_{\mathrm{rel}}$ & median & max \\
    \midrule
    ORB-v3-consv-inf-OMAT & 364 & \textbf{452} & 18 & \textbf{424} & 3\,843 \\
    MACE-MH-1-OMAT & 475 & \underline{477} & 19 & \underline{441} & \textbf{2\,726} \\
    MatterSim-5M & 580 & 552 & 22 & 506 & \underline{3\,566} \\
    MACE-MP-0b3 & \underline{294} & 809 & 32 & 705 & 6\,240 \\
    MACE-OMAT-0 & \textbf{233} & 645 & 25 & 569 & 9\,979 \\
    UMA-S-1.1-OMAT & 3\,015 & 10\,187 & 401 & 4\,054 & 181\,015 \\
    \bottomrule
  \end{tabular}
\end{table}

\begin{table}[htbp]
    \caption{\textbf{Energy and force errors for general bulk structures.}
    MAEs and per-structure force-error statistics over 5300 general bulk
    structures. Of these structures, 1414 have
    reference forces that vanish by symmetry. Energy and force units are meV\,atom$^{-1}$ and meV\,\AA$^{-1}$, respectively. Bold and underlined
    values mark the best and second-best results.}
  \label{tab:errors_bulk}
  \centering
  \footnotesize
  \setlength{\tabcolsep}{5pt}
  \begin{tabular}{l ccc cc}
    \toprule
    \multirow{2}{*}{Foundation MLIP} & \multicolumn{3}{c}{MAE}
      & \multicolumn{2}{c}{Per-structure force MAE} \\
    \cmidrule(lr){2-4}\cmidrule(lr){5-6}
     & $E$ & $F$ & $F_{\mathrm{rel}}$ & median & max \\
    \midrule
    ORB-v3-consv-inf-OMAT & \textbf{182} & \textbf{426} & 26 & \textbf{79} & \textbf{1\,592} \\
    MACE-MH-1-OMAT & 755 & 18\,998 & 1164 & 122 & 5\,073\,289 \\
    MatterSim-5M & \underline{183} & \underline{640} & 39 & 106 & \underline{2\,492} \\
    MACE-MP-0b3 & 559 & 1\,336 & 82 & 277 & 12\,954 \\
    MACE-OMAT-0 & 589 & 15\,695 & 962 & \underline{90} & 573\,612 \\
    UMA-S-1.1-OMAT & 3\,900 & 1\,081 & 66 & 244 & 49\,096 \\
    \bottomrule
  \end{tabular}
\end{table}

\clearpage
\section{Binding curves}
\label{app:physicality}

\begin{figure}[htbp]
  \centering
  \includegraphics[width=\linewidth]{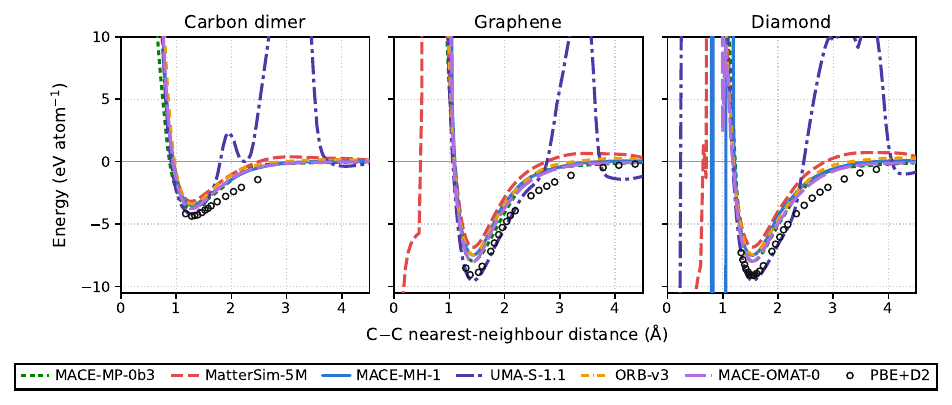}
  \caption{\textbf{Binding curves with D3 dispersion.}
  Energy per atom against nearest-neighbour C--C distance for the carbon dimer
  and isotropically scaled graphene and diamond cells. D3 is applied to each
  model, matching the molecular-dynamics protocol, while the reference remains
  PBE+D2~\cite{qamar2023atomic}. Relative to the D2-corrected curves in
  Figure~\ref{fig:binding_D2}, energies differ by at most
  0.1~eV\,atom$^{-1}$ over the displayed range.}
  \label{fig:binding_D3}
\end{figure}

\begin{figure}[htbp]
  \centering
  \includegraphics[width=\linewidth]{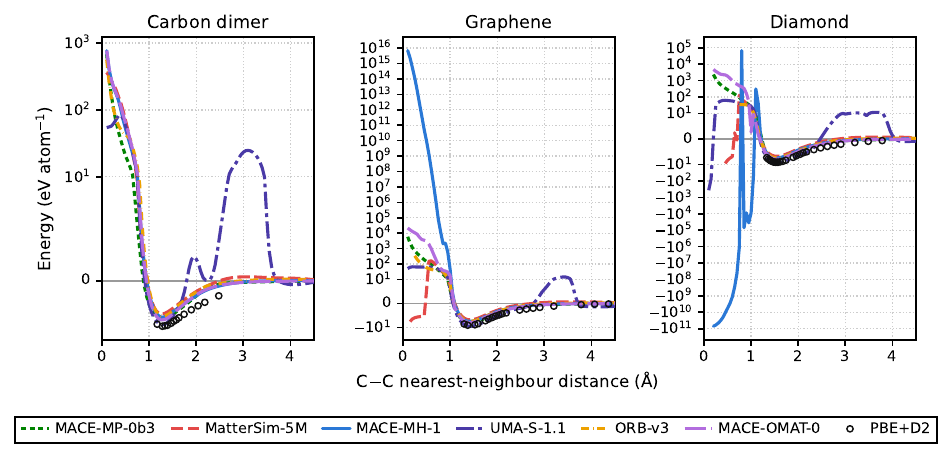}
  \caption{\textbf{Full energy range of the D2-corrected binding curves.}
  The curves from Figure~\ref{fig:binding_D2} are shown without the
  $\pm10$~eV\,atom$^{-1}$ y-limits, using a logarithmic axis.
  These curves show the full barriers and compression-induced PES
  holes.}
  \label{fig:binding_log}
\end{figure}

The main-text binding curves use D2 dispersion to match the PBE+D2 reference data, whereas D3 is applied during molecular dynamics. Figure~\ref{fig:binding_D3} shows that replacing D2 with D3 changes the model energies by at most 0.1~eV\,atom$^{-1}$ over the displayed range and does not alter the qualitative conclusions.
Figure~\ref{fig:binding_log} shows the same D2-corrected curves over their full energy range, exposing the barriers and compression-induced PES holes discussed in the main text.

\clearpage
\section{Generation protocol}
\label{app:protocol}

\begin{figure}[htbp]
  \centering
  \includegraphics[width=\linewidth]{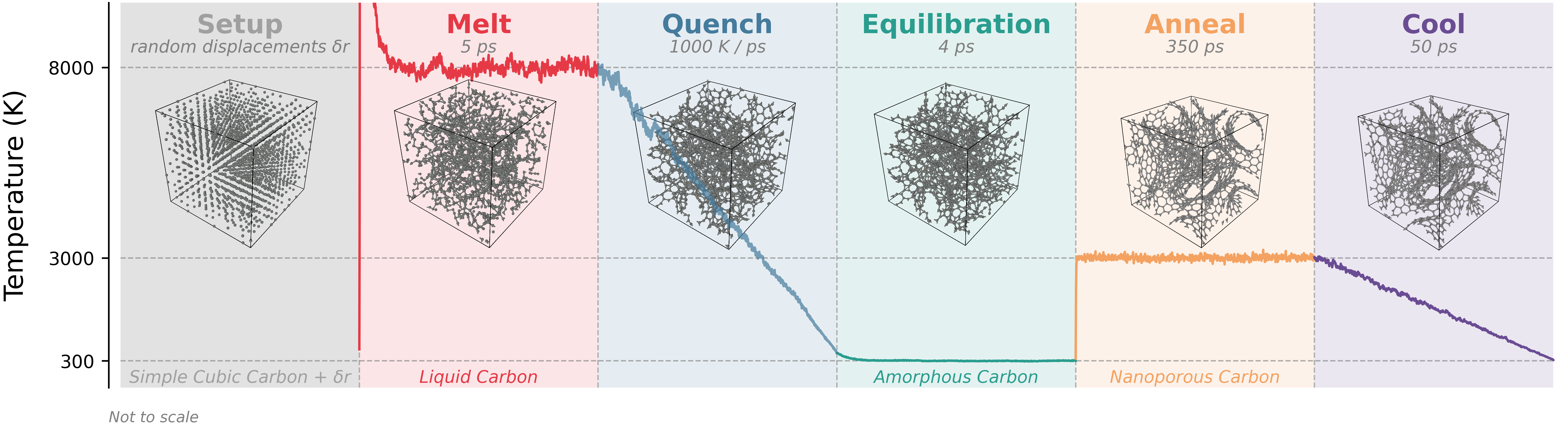}
  \caption{\textbf{Melt-quench-anneal-cool generation protocol.} Temperature against simulation
  time through the six stages, with a representative snapshot for each. A simple-cubic lattice with
  random displacements is melted at 8,000~K, quenched to 300~K at 1,000~K\,ps$^{-1}$, equilibrated,
  annealed at 3,000~K to promote graphitisation, then cooled to 300~K. The time axis is not to
  scale.}
  \label{fig:protocol}
\end{figure}

Figure~\ref{fig:protocol} shows the generation protocol used for the MD stability tests on 216 atoms, as well as the larger 1728-atom simulations.
Initial structures were constructed from a simple-cubic unit cell containing one carbon atom. The unit cell was replicated along each axis to obtain the desired supercell and system size, after which the cell dimensions were scaled to the target mass density. Small random displacements were applied to break the initial symmetry. For O/H-containing systems, randomly selected carbon sites were replaced with oxygen or hydrogen to approximate the composition of the commercial nanoporous carbon YP-50F (93.5:1.1:5.5 C:O:H).
The initial structures were melted at 8,000~K for 5~ps and quenched to 300~K at a rate of 1,000~K\,ps$^{-1}$. Following equilibration at 300~K for 4~ps, the systems were annealed at 3,000~K for 350~ps to promote graphitisation. Finally, they were cooled from 3,000 to 300~K over 50~ps to approximate experimental cooling and encourage the formation of OH-including functional groups observed in experimental carbons. A timestep of 0.5~fs was used for the melt and quench, before switching to 1~fs for the 350~ps anneal.
MACE simulations were performed using \texttt{symmetrix}~\cite{kovacs2025mace,batatia2025foundation} in LAMMPS~\cite{thompson2022lammps}. ORB simulations were performed using ASE~\cite{ase-paper}. Double-precision floating-point arithmetic (\texttt{float64}) was used throughout to improve numerical stability, particularly during the high-temperature stages. A D3 dispersion correction~\cite{grimme2010consistent} was applied consistently to all models, with the accelerated \texttt{AlchemiDFTD3} implementation used for ORB. Temperature was controlled using a canonical-sampling-through-velocity-rescaling thermostat~\cite{bussi2007canonical}.

\section{Coordination number}

\begin{figure}[htbp]
  \centering
  \includegraphics[width=\linewidth]{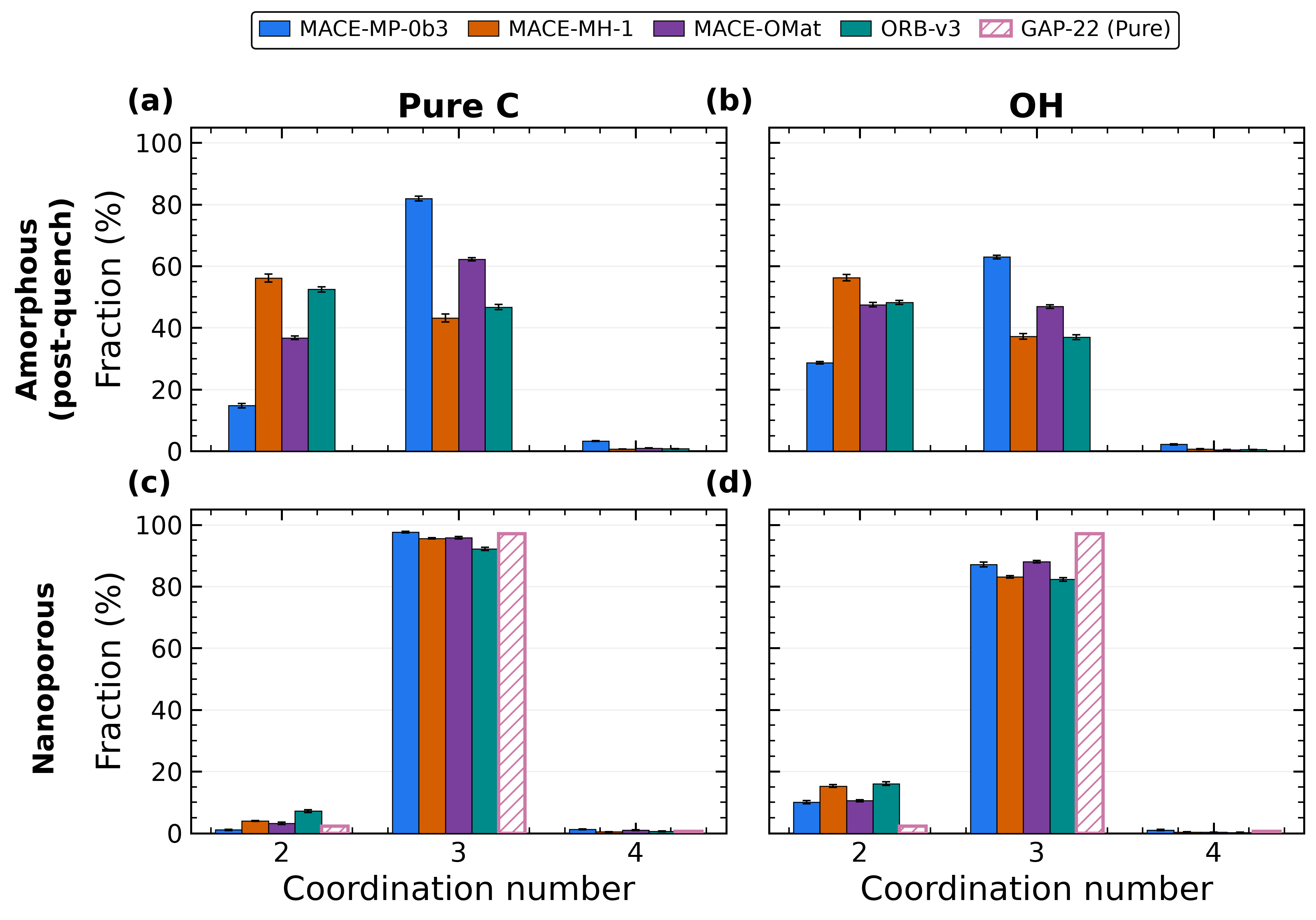}
  \caption{\textbf{Coordination number for pure and O/H-containing amorphous and nanoporous carbons.} Each bar represents the fraction of carbon atoms which are connected to $N$ other atoms. Nanoporous carbons are compared to a pure-carbon GAP-22 reference~\cite{wang2022structure}.}
  \label{fig:app_coord}
\end{figure}

Figure~\ref{fig:app_coord} shows that the amorphous-carbon precursors generated by the
OMAT-based models (ORB-v3, MACE-OMAT-0 and MACE-MH-1) contain substantially more
under-coordinated carbon than the MACE-MP-0b3 precursor. During annealing, local bond closure at
these sites promotes five- and seven-membered rings, introducing curvature and reducing graphitic
order. The difference between model families is smaller for O/H-containing carbon because oxygen
and hydrogen can passivate under-coordinated sites.

\section{Pore size convergence}
\label{app:PSD}

\begin{figure}[htbp]
  \centering
  \includegraphics[width=0.6\linewidth]{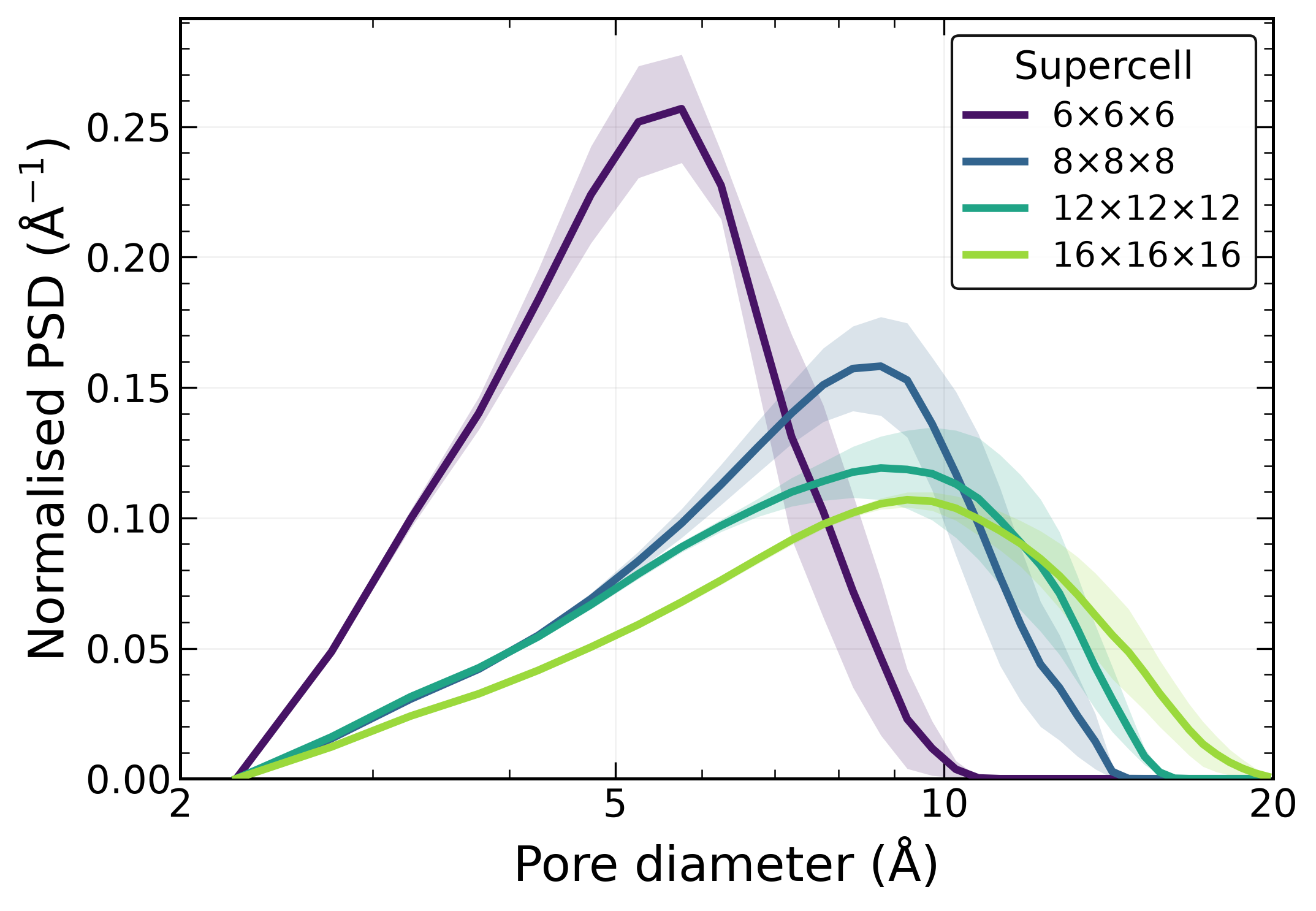}
  \caption{\textbf{Characteristic pore size is converged by $12\times12\times12$.} Pore-size distributions of MACE-MP-0b3 structures generated at 1.0~g\,cm$^{-3}$ using
  supercells from $6\times6\times6$ to $16\times16\times16$. Each curve is averaged over five
  independent simulations and evaluated from the final cooled and relaxed configurations.}
  \label{fig:app_PSD}
\end{figure}

Figure~\ref{fig:app_PSD} examines the dependence of the calculated pore-size distribution on simulation-cell size. 
Increasing the supercell from $6\times6\times6$ to $8\times8\times8$ substantially reduces the peak amplitude and shifts the distribution towards larger pore diameters.
Small simulation cells restrict the maximum pore size that can be accommodated and contain fewer independent pores, allowing a single void to dominate the distribution.
The peak position for the $8\times8\times8$ supercell broadly agrees with those obtained using the $12\times12\times12$ and $16\times16\times16$ cells, although the maximum shifts slightly towards larger diameters and decreases in amplitude with increasing system size. 
The difference between the $12\times12\times12$ and $16\times16\times16$ results is comparatively small and is primarily associated with a longer tail towards large pore sizes in the larger cell.
The $12\times12\times12$ cells used in the main analysis therefore provide an adequately converged characteristic pore size, although the peak amplitude and large-pore tail retain some finite-size dependence.

\section{Software released with this work \& data availability}
\label{app:software}

We developed \texttt{PoreTorch}, a GPU-accelerated Python package for geometric pore analysis. The PSDs in Figure~\ref{fig:validation} use local-clearance diameters weighted by the corresponding sphere volumes, with each run normalised before averaging. Batched \texttt{PyTorch} operations accelerate the grid-based calculation. A benchmark of the original local-clearance implementation on a 1728-atom structure with \SI{0.5}{\angstrom} grid spacing achieved approximately $230\times$ acceleration on an NVIDIA GeForce RTX 5090 relative to a single-process \texttt{NumPy} baseline. Implementation details and documentation are available at \url{https://github.com/joehart2001/poretorch}.

Pair distribution functions are calculated using \texttt{cuRDF} (\url{https://github.com/joehart2001/cuRDF}), which uses NVIDIA ALCHEMI Toolkit-Ops neighbour lists for GPU accelerated RDF calculations.

The accompanying analysis and data repository provides the structures, numerical results and analysis functions for the structural validation in Figure~\ref{fig:validation}, including the calculation parameters. It is available at \url{https://github.com/joehart2001/neurips2026_sim2science_carbon}.

\clearpage
\section{Compute and assets}
\label{app:compute}

\paragraph{Compute.}
MACE trajectories were run with LAMMPS~\cite{thompson2022lammps} on eight ARCHER2 CPU nodes, using 1024 single-threaded MPI ranks. Typical wall times were approximately 4.5~h for 216 atoms and 10~h for 1728 atoms. ORB trajectories used a single NVIDIA A100 40~GB GPU through ASE~\cite{ase-paper} with \texttt{AlchemiDFTD3}, requiring 5--8~h for 216 atoms and 1-2 days for 1728 atoms.

\paragraph{Assets.}
Models were evaluated as released, without retraining. Table~\ref{tab:assets} lists the model checkpoints, reference data and principal software used. The public analysis packages and structural-validation data are linked in Appendix~\ref{app:software}. \texttt{PoreTorch} and \texttt{cuRDF} are released under the MIT licence. The data repository covers Figure~\ref{fig:validation}. It does not include a complete archived environment or all benchmark and molecular-dynamics workflows.

\begin{table}[htbp]
  \caption{Principal assets used in this work. Model licences apply to the released weights.}
  \label{tab:assets}
  \centering
  \footnotesize
  \setlength{\tabcolsep}{3pt}
  \begin{tabular}{@{}
    p{0.28\linewidth}
    p{0.19\linewidth}
    p{0.28\linewidth}
    p{0.17\linewidth}
    @{}}
    \toprule
    Asset & Type & Release or checkpoint & Licence \\
    \midrule

    \multicolumn{4}{@{}l}{\textit{Model weights}} \\

    MACE-MP-0b3~\cite{batatia2025foundation}
      & Foundation MLIP
      & 0b3, MACE 0.3.9+
      & MIT \\

    MACE-OMAT-0~\cite{batatia2025foundation}
      & Foundation MLIP
      & 0, MACE 0.3.10+
      & ASL v1.0 \\

    MACE-MH-1-OMAT~\cite{batatia2025cross}
      & Foundation MLIP
      & 1, \texttt{omat\_pbe} head, MACE 0.3.14+
      & ASL v1.0 \\

    MatterSim-5M~\cite{yang2024mattersim}
      & Foundation MLIP
      & v1.0.0-5M
      & MIT \\

    ORB-v3-consv-inf-OMAT~\cite{rhodes2025orb}
      & Foundation MLIP
      & v3 conservative-inf-OMAT
      & Apache-2.0 \\

    UMA-S-1.1-OMAT~\cite{wood2025family}
      & Foundation MLIP
      & S-1.1, OMAT task
      & FAIR Chemistry License v1 \\

    GAP-22~\cite{wang2022structure}
      & Comparison potential
      & v2.0
      & CC BY 4.0 \\

    \addlinespace
    \multicolumn{4}{@{}l}{\textit{Reference data}} \\

    Diverse carbon~\cite{qamar2023atomic}
      & Reference DFT dataset
      & Figshare file \texttt{41339219}
      & CC BY-NC 4.0 \\

    Graphene oxide~\cite{el2024accelerated}
      & Reference DFT dataset
      & Zenodo v2, record \texttt{14066557}
      & CC BY 4.0 \\

    Ti-CDC-\SI{1000}~\cite{forse2015new}
      & Experimental PDF and PSD
      & 2015 publication
      & Not separately stated \\

    \addlinespace
    \multicolumn{4}{@{}l}{\textit{Software}} \\

    LAMMPS~\cite{thompson2022lammps}
      & MD engine
      & Build not specified
      & GPL-2.0-only \\

    \texttt{symmetrix}~\cite{kovacs2025mace,batatia2025foundation}
      & MACE--LAMMPS interface
      & Build not specified
      & MIT, interface GPL-2.0 \\

    ASE~\cite{ase-paper}
      & MD driver and I/O
      & Build not specified
      & GNU LGPL \\

    \texttt{PoreTorch}
      & Pore analysis
      & 0.1.0, sphere-weighting update
      & MIT \\

    \texttt{cuRDF}
      & Pair distributions
      & 0.5.4
      & MIT \\

    \bottomrule
  \end{tabular}
\end{table}

\end{document}